\RequirePackage{lineno}
\documentclass[aps,prb,showpacs, groupedaddress,twocolumn,superscriptaddress]{revtex4}
\usepackage{amsmath,graphicx,latexsym,times,color}
\usepackage{setspace}
\usepackage{hyperref}
\usepackage{array}
\usepackage{titlesec}
\usepackage{physics}
\usepackage{xcolor}

\let\oldtimes\times  
\renewcommand\times{{\oldtimes}}

\begin{document}
	
    \title{Proposal for all-electrical spin manipulation and detection for a single molecule on boron-substituted graphene}

	\author{Fei Gao}
	\affiliation{Department of Physics, Technical University of Denmark, DK-2800 Kongens Lyngby, Denmark}
	
	\author{Dongzhe Li}
	\email{dongzhe.li@cemes.fr}
	\affiliation{CEMES, Universit\'e de Toulouse, CNRS, 29 rue Jeanne Marvig, F-31055 Toulouse, France}
	
	\author{Cyrille Barreteau}
	\affiliation{DRF-Service de Physique de l’Etat Condens\'e, CEA-CNRS, Universit\'e Paris-Saclay, F-91191 Gif-sur-Yvette, France}
	
	\author{Mads Brandbyge}
	\email{mabr@dtu.dk}
	\affiliation{Department of Physics, Technical University of Denmark, DK-2800 Kongens Lyngby, Denmark}
	\affiliation{Center for Nanostructured Graphene, Technical University of Denmark, DK-2800 Kongens Lyngby, Denmark}

	\date{\today}
	
	\begin{abstract}
		All-electrical writing and reading of spin states attract considerable attention for their promising applications in energy-efficient spintronics devices. Here we show, based on rigorous first-principles calculations, that the spin properties can be manipulated and detected in molecular spinterfaces, where an iron tetraphenyl porphyrin (FeTPP) molecule is deposited on boron-substituted graphene (B-G). Notably, a reversible spin switching between the $S=1$ and $S=3/2$ states is achieved by a gate electrode. We can trace the origin to a strong hybridization between the Fe-$d_{{z}^2}$ and B-$p_z$ orbitals. Combining density functional theory with nonequilibrium Green's function formalism, we propose an experimentally feasible 3-terminal setup to probe the spin state. Furthermore, we show how the in-plane quantum transport for the B-G, which is non-spin polarized, can be modified by FeTPP, yielding a significant transport spin polarization near the Fermi energy ($>10\%$ for typical coverage). Our work paves the way to realize all-electrical spintronics devices using molecular spinterfaces.	
	\end{abstract}
	
	\renewcommand{\vec}[1]{\mathbf{#1}}
	
	\maketitle
	
	Achieving size-compact and energy-efficient control and detection of magnetism are paramount for the development of future spintronic devices. Using single molecules as quantum units opens a new pathway to reach the physical limits of miniaturization. Currently, spintronics devices are mainly operated via either an external magnetic field (e.g., tunnel magnetoresistance devices \cite{Butler2001}) or electric currents (e.g., spin-transfer torque devices \cite{parkin2008magnetic}), which are both highly power-consuming. More recently, electric-field manipulation of magnetism has been proposed \cite{matsukura2015control,song2017recent} and has been extensively studied in bulk and 2D materials \cite{mak2019probing,forces3}. However, the full-electrical programmable reading and writing of magnetism at the single-molecule level are still unsolved problems.

	It is now well known that in molecular spintronincs most of the phenomena are driven by the interface, which leads to the concept of spinterface\cite{barraudUnravellingRoleInterface2010,sanvitoRiseSpinterfaceScience2010,cinchetti2017}. Ideally, one aims at the control and detection\cite{schedin2007detection,holovchenko2016near} at the individual molecule limit. Therefore, single molecules asorbed on surfaces have become an ideal testbed to study the interaction of molecules with surfaces, the surrounding environment, and responses to external chemical stimuli. In particular, controlling molecular spin states by chemical functionalization of the surface allows for creating molecular devices with novel functionalities \cite{2021spin}. During the last decade, particular attention has been focused on the substitution of carbon atoms in the graphene lattice by heteroatoms leading to new physical and chemical properties \cite{panchakarla2009synthesis,Wang2012,Sforzini2016,agnoli2016doping}. In nitrogen-substituted graphene (N-G), scanning tunneling microscopy (STM)\cite{Tison2015} showed a dramatic change of the local electronic structure around the nitrogen depending on its surroundings. This suggests that the N-G may be used to tune the properties of adsorbed molecules, and indeed, the adsorption on top of the N site of N-G can modify the molecular levels by shifts\cite{Pham2014}, charge transfer and level splitting\cite{Npair}, or change of spin state\cite{de2018non}. 
	
	
	Alternatively, boron (B) is also suitable for direct incorporation into the graphene honeycomb lattice, resulting in at least an effective $p$-doping  \cite{zhao2013local,Gebhardt2013,pan2015room,kawai2015atomically,usachov2016large}. However, unlike N-G, the B-substituted graphene-based molecular interfaces have been much less investigated in both theory and experiment. Therefore, detailed insight into the interaction between molecule and B-graphene (B-G) at the atomic scale is currently lacking.
	
	In this Letter, we propose the B-G substrate as an ideal spinterface for molecular magnets: We demonstrate, using density functional theory (DFT), the {\em electrical tuning and probing of spin states} in a single-molecule device adsorbed on B-G. We choose iron tetraphenyl porphyrin (FeTPP), which has different magnetic ground states on graphene and Au surfaces\cite{rubio2018orbital, FeTPP_Gr}. We find that a single FeTPP molecule on B-G allows for a reversible spin transition between $S=1$ and $S=3/2$ controlled by an external electrical gate. This effect is driven by a strong and tunable hybridization between FeTPP and B-G. Combining DFT with Keldysh Green's function techniques, we further propose an experimentally feasible 3-terminal transport setup to probe the transport spin polarization (TSP). In contrast to pristine and N-G substrates, the in-plane spin transport for the B-G is significantly modified by the FeTPP with a TSP of more than $10\%$ for a typical coverage. Our work shows a promising application to all-electrically writing and reading magnetization states in molecular spintronics devices.	
	\begin{figure}[!t]
		\centering
		\includegraphics[width=0.68\linewidth]{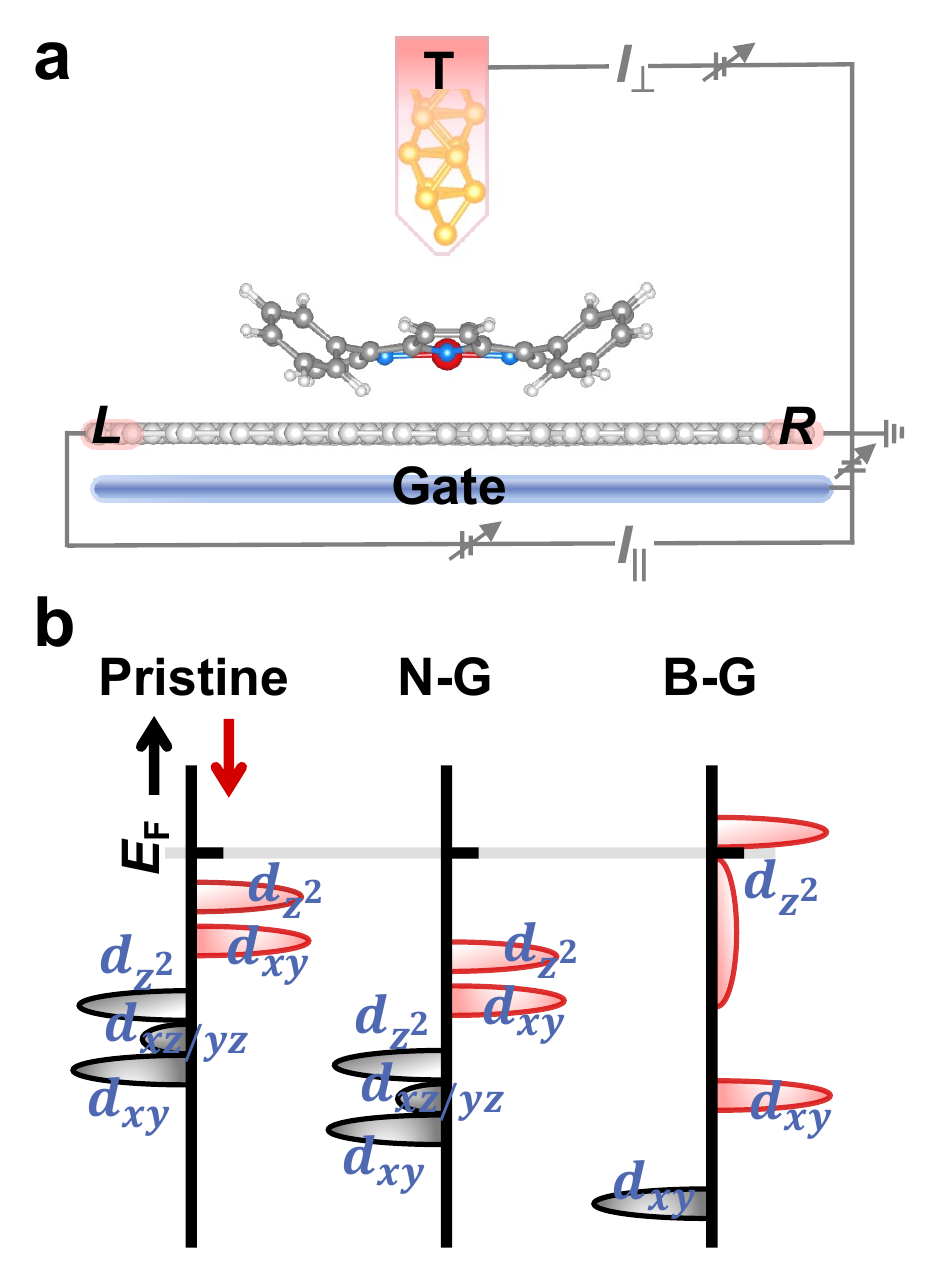}
		\caption{\label{fig:geometry}
			(a) Schematic view of a single FeTPP absorbed in substituted graphene with a model STM tip and a charge back-gate plane. The proposed 3-terminal device allows an out-of-plane ($I_{\perp}$) and an in-plane transport ($I_{\parallel}$). (b) Sketches of PDOS on Fe $d$-orbitals for the molecule deposited on pristine, N-substituted, and B-substituted graphene.}
	\end{figure}
	
The transport setup (see Fig. \ref{fig:geometry}a.) consists of graphene with two contacts (L,R) and a charge-plane mimicking the back-gate underneath, and an Au tip electrode above the FeTPP.  This setup allows for two possible current flows: The in-plane transport from left (L) to the right (R) graphene electrodes ($I_{\parallel}$), and the out-of-plane from L/R to Tip ($I_{\perp}$). This is feasible in state-of-the-art STM\cite{li2018survival,Jingcheng_2019}, where the back-gate charge is capacitively controlled by a gate voltage. The gate charge enables ``writing", while the $I_{\parallel}$ or $I_{\perp}$ currents  provide a ``reading" of the FeTPP spin state.
The electronic structure calculations were performed using \textsc{SIESTA} \cite{brandbyge2002density} within the $\text{DFT}+U$ scheme, and checked by comparing to plane-wave calculations \cite{vasp}. The transport was studied using \textsc{TransSIESTA} \cite{brandbyge2002density, siesta, papior2017improvements} code, which employs the non-equilibrium Green’s function (NEGF) formalism combined with DFT, and “post-processing” codes \textsc{Tbtrans} and \textsc{SISL} \cite{sisl}. We refer to Supplemental Material \MakeUppercase{\expandafter{\romannumeral1}} \cite{comment} for computational details.
	
	
	\begin{figure}[!t]
		\centering
		\includegraphics[width=1\linewidth]{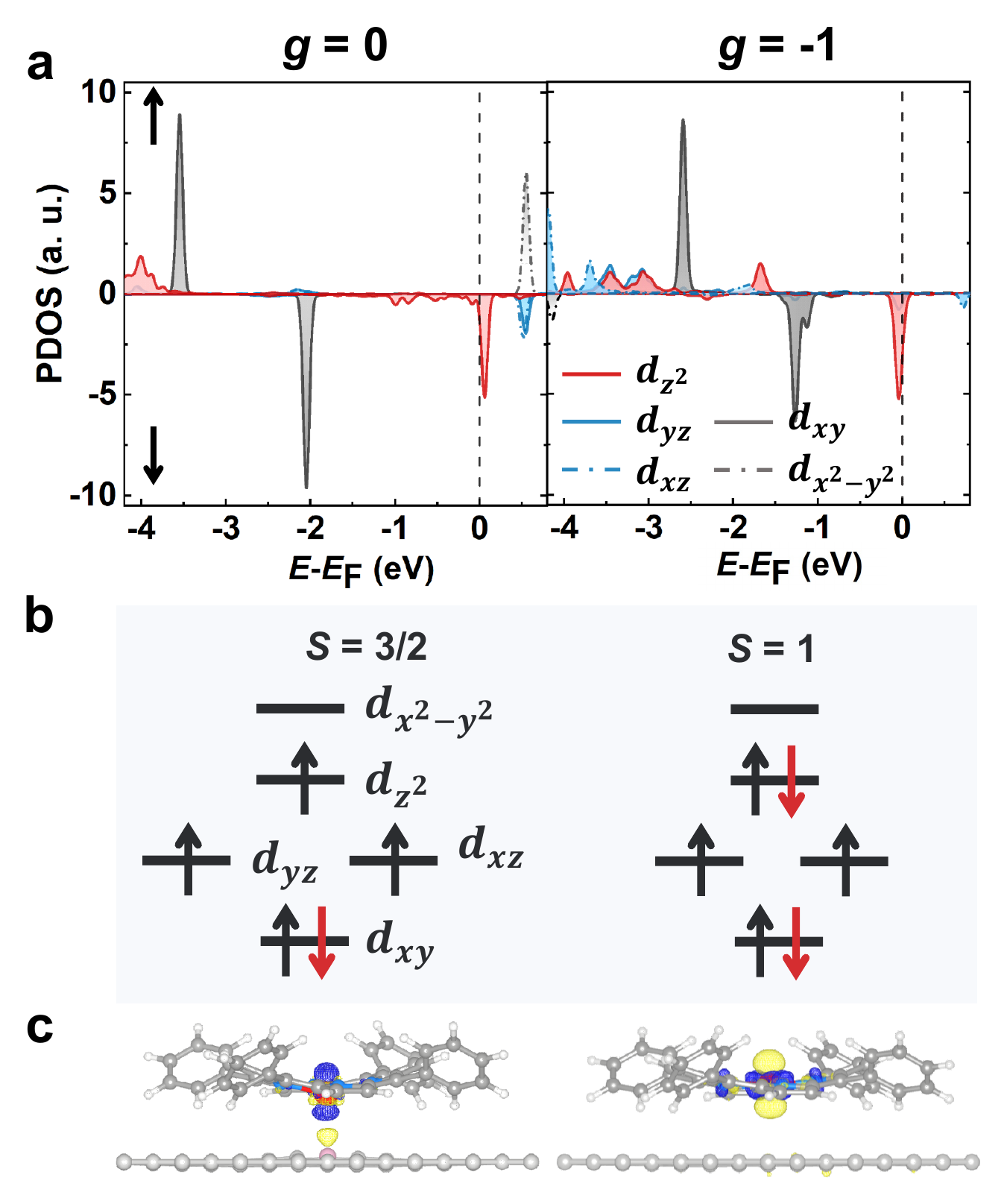}
		\caption{\label{pdos} Electrical writing of spin states for a single FeTPP absorbed on B-substituted graphene. (a) Calculated PDOS on Fe $d$-orbitals with $\emph{g}=0$ and $\emph{g}=-1$. (b) Occupation of Fe $d$-orbitals in the $S=3/2$ and $S=1$ states. In particular, an occupation transition (from unoccupied to occupied state) is observed for $d_{z^2}$ orbital when the gate charge is applied. (c) Charge density difference for $\emph{g}=0$ ($\Delta \rho = \rho_{\text{FeTPP+B-G}}-\rho_{\text{FeTPP}}-\rho_{\text{B-G}}$) and $\emph{g}=-1$ ($\Delta \rho = \rho_{\text{\emph{g}=-1}}-\rho_{\text{\emph{g}=0}}$). Yellow and blue clouds correspond to electron accumulation and depletion, respectively. Isosurface values of $\pm 0.005$ e/$\text{bohr}^\text{3}$.}	\end{figure}
	
	The electronic properties of the isolated FeTPP have been thoroughly explored both experimentally and theoretically, revealing the dependence of the spectroscopic state on the environment of the Fe atom that gives a ground state either in low ($S=0$), high ($S=2$), or an intermediate spin state ($S=1$). In general, the ground states of the free FeTPP is $S=1$ state with $^3\!A_{2g}$ having the occupancy of the 3$d$ shell  $(d_{xy})^2(d_{z^2})(d_{xz})^1(d_{yz})^1$, the last two orbitals being degenerate as a consequence of the molecular symmetry \cite{liao2002electronic, FeTPP_Gr}. Figure \ref{fig:geometry}b shows schematic sketches of spin-polarized projected densities of states (PDOS) of the FeTPP on doped graphene (without the STM tip). It has been demonstrated through combined STM experiments and DFT calculations that the molecule keeps the same electronic structure as in the gas phase after being deposited on pristine graphene due to weak molecule-substrate coupling, and the HOMO state originates essentially from the Fe $d_{z^2}$ spin-down orbital \cite{FeTPP_Gr}. In the case of N-G, the molecule remains $S = 1$ with the same spin configuration, while a clear downshift of the electronic spectrum is observed in good agreement with experiments\cite{FeTPP_Gr}. Surprisingly, a significant change happens when the molecule is attached to B-G: the total magnetic moment varies from 2 to 3 $\mu_\text{B}$ with a change of oxidation state from Fe$^{2+}$ to Fe$^{3+}$. This variation is due to a chemical absorption nature as reflected in the calculated binding energy of 2.9 eV. On pristine graphene and N-G we tested different FeTPP orientations and positions, such as hollow and bridge site, showing little difference in electronic and magnetic properties. Meanwhile, the strong coupling between FeTPP and B-G leads to a clear preference for the Fe atom on top of the B atom at a distance ($d_{\text{Fe-Gr}}$) of 2.7~{\AA}. The B-G leads to a redistribution of electron density, where the electron-deficient boron sites provide the enhanced binding capability\cite{B-doping}. The Mulliken charge analysis shows that upon adsorption 0.7 electrons are transferred from Fe to B-G. In particular, the spin-down channel of Fe-$d_{z^2}$, which was initially occupied, becomes an empty state just above $E_{\text{F}}$ due to strong hybridization between Fe-$d_{z^2}$ and B-$p_z$ by perfect orbital symmetry matching, as shown in Fig. \ref{pdos}a. Such strong hybridization is also reflected in the charge density difference plotted in Fig. \ref{pdos}c.
	Furthermore, the strong interaction also introduces a small spin-polarization of the B atom and the carbon atoms at the spinterface. For more details, See Supplemental Material \MakeUppercase{\expandafter{\romannumeral2}} \cite{comment}.

	
	\begin{figure}[!t]
		\centering
		\includegraphics[width=0.9\linewidth]{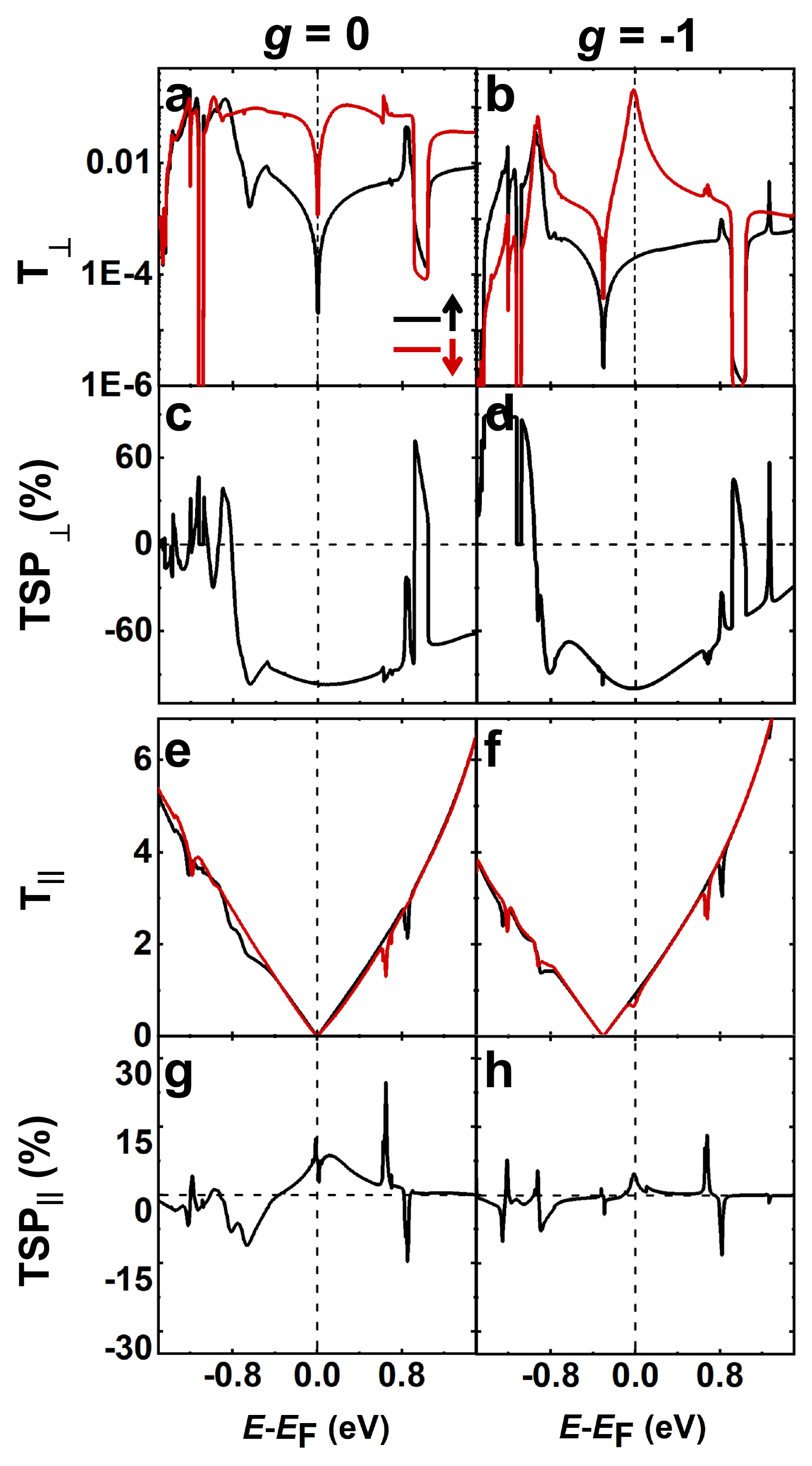}
		\caption{\label{T-mole-B-G} Electrical readout of spin states for a single FeTPP absorbed on B-substituted graphene. (a-b) Out-of-plane spin-dependent transmission functions with $\emph{g}=0$ and $\emph{g}=-1$. Pink and blue lines denote spin-up and spin-down channels, respectively. (c-d) The out-of-plane transport spin polarization (e-f) In-plane spin-dependent transmission with $\emph{g}=0$ and $\emph{g}=-1$. (g-h) The in-plane transport spin polarization. (TSP defined as $\text{TSP} = T_{\uparrow}-T_{\downarrow})/(T_{\uparrow}+T_{\downarrow}$) for $\emph{g}=0$ and $\emph{g}=-1$.}
	\end{figure}

	
	A method to achieve full-electrical writing of the magnetization states at the single-molecule level is attractive and promising, corresponding to a reversible spin manipulation. The coupling between FeTPP and B-G at the interface leads to an empty $d_{z^2}$ orbital very close to the Fermi energy. It turns out that external stimuli may easily tune it: We apply a gate plane placed 15~{\AA} underneath the graphene, as shown in Fig. \ref{fig:geometry}a. The gate carries a charge density of $n=\emph{g}\times 10^{13} e/\text{cm}^2$, where $\emph{g}$ defines the gating level, with $\emph{g}<0$ ($\emph{g}>0$) corresponding to $\emph{n}$ ($\emph{p}$)-type doping \cite{gating}. Here, the FeTPP molecule and the second nearest-neighbor C atoms to the B atom were fully relaxed when applying a gate charge. Figure \ref{pdos}a shows the corresponding DOS projected onto Fe $d$-orbitals for $\emph{g}=0$ and $\emph{g}=-1$. The variation of PDOS for both spin channels is significant. The $d_{xy}$ orbitals shift to higher energy with $\emph{g}=-1$, although it has no direct influence on the total magnetic moment. In contrast, the $d_{z^2}$ spin down becomes occupied, leading to a spin switching from $S=3/2$ to $S=1$, also as shown in Fig. \ref{pdos}b. The real-space charge density difference in Fig. \ref{pdos}c also indicates that the molecule retrieves the $d_{z^2}$ electron when $\emph{g}=-1$, compared to $g=0$. The atomic structures in Fig. \ref{pdos}c clearly show the bond elongation/weakening between the relaxed molecule and B-G with $\emph{g}=-1$. For the unrelaxed, ungated structure subject to the $\emph{g}=-1$ gate, the gate-induced forces increasing the Fe-B distance are substantial, 0.2 nN and 0.6 nN on the B and Fe atom, respectively, pushing the structure towards the weakly bonded situation found for pristine graphene. 
	For completeness, we note that for the weak coupling configuration at $\emph{g}=-1$ we also get a competing $S=1$ solution, $(d_{xy})^2(d_{z})^1(d_{xz})^2(d_{yz})^1$, only 10~meV higher in energy, see Supplemental Material \MakeUppercase{\expandafter{\romannumeral2}} \cite{comment}.
	
	In order to understand the gate-control of the spin states of the FeTPP molecule, we have performed test calculations with $\emph{g}=-1$ for B-G without FeTPP. When $\emph{g}=-1$, the B-G substrate becomes charged with extra electrons and thus loses its ability to attract the FeTPP. In other words, the electronic spin writing in the molecular spinterface can be achieved by a gate due to the tunability of the interaction between FeTPP and B-G.
	Increasing the gate charge to $\emph{g}=-2$ yield little difference compared to $\emph{g}=-1$, while the spin state is not tunable with $\emph{g}>0$ where it remains $S=3/2$. Although the case of FeTPP may seem specific for B-G, the proposed mechanism is quite generic, and a similar approach should be possible when the frontier molecular orbital is close to $E_{\text{F}}$.
	
	
	
	
	For the electrical read-out of the single molecular spin states, we first consider the out-of-plane spin transport from the L/R graphene electrode to the Au tip electrode. Figure \ref{T-mole-B-G}a and b show the spin-dependent transmission functions with $g=0$ and $g=-1$. We observe the transmissions exhibit a dip near the Fermi level due to the vanishing DOS in graphene at $g=0$, and it shifts to below $E_{\text{F}}$ when $g=-1$. We find almost fully spin-polarized current near $E_{\text{F}}$ for both $g=0$ and $g=-1$, as shown in Fig. \ref{T-mole-B-G}c and d.  The scanning tunneling spectroscopy ($dI/dV$ curve), probing the energy dependence of $T_\perp$, should easily distinguish the two different spin states at $g=0$ and $g=-1$, while shot noise measurements \cite{Mohr-noise-19,gehring2019single}, yielding TSP$_{\perp}$, would not. The STM tip may furthermore be used actively to manipulate the molecules and control their spin\cite{tip1}.	
	%
	
	
	
	Next, we consider the in-plane transport where the electrical current runs through the graphene $xy$ plane (from L to R). Since we use pristine graphene as $\text{L/R}$-electrodes, the Fermi energy is positioned at the Dirac point for $g=0$. Without FeTPP, the N-G and B-G systems retain the pristine sp$^2$ hybridization and conjugated planar structure, leading to a non-spin-polarized behavior. This is in contrast to the spin-polarized current reported in the case of B-substituted graphene nanoribbons \cite{Martins2007}. To understand how the transport properties of the substrates are modified through the molecular interfacial hybridization effects, we plot in Fig. \ref{T-mole-B-G}e-f the corresponding transmission functions with FeTPP adsorbed on B-G at $g=0$ and $g=-1$. Interestingly, our calculations show that when $g=0$, the spin-up and spin-down molecular orbitals hybridize very differently with the substrate, resulting in clear spin-dependent transport behavior (Fig. \ref{T-mole-B-G}g). For the spin-down channel (red lines), the transmission becomes almost linear at $E < E_{\text{F}}$ (similar to pristine B-G), which is significantly different from the rather broadened feature without FeTPP. On the other hand, the transmission for the spin-up (black lines) resembles B-G transport without molecule. As a result, we get a large transport spin polarization (TSP$_{\parallel}$), defined as $\text{TSP}_{\parallel} = (T_{\parallel}^{\uparrow}-T_{\parallel}^{\downarrow})/(T_{\parallel}^{\uparrow}+T_{\parallel}^{\downarrow})$, effect near the Fermi energy where it furthermore changes sign. For the applied $y$-periodic transport cell the $\text{TSP}_{\parallel}$ reach values beyond $10\%$ which is significant when we consider the corresponding inter-molecular distance of $\sim16$~{\AA}, below typical coverages\cite{FeTPP_Gr}.
	
	
	On the other hand, the spin-up and spin-down transmission functions are almost degenerate when $g=-1$, resulting in the absence of $\text{TSP}_{\parallel}$ (Fig. \ref{T-mole-B-G}h). Such an on-off TSP via gating could be probed, e.g., in shot noise experiments \cite{Mohr-noise-19,gehring2019single}. It should be noted that the concentration of FeTPP in typical experiments will be higher than our quasi-single molecule setup, which further increases the TSP. In addition, for pristine and N-G, the transmission remains almost the same as without FeTPP due to weak coupling between molecule and substrate (i.e., physisorption), leading to non-spin-polarized current (See Supplemental Material \MakeUppercase{\expandafter{\romannumeral3}} \cite{comment}.).

	

	
	In summary, we propose a molecular spinterface device based on FeTPP on a B-substituted graphene substrate. Our calculations demonstrate all-electrical writing and reading of magnetization states at the single-molecule level. The spin states of FeTPP can be switched reversely between $S=3/2$ and $S=1$, tracing the origin to the strong hybridization between Fe-$d_{z^2}$ and B-$p_z$ orbitals. We further propose a 3-terminal transport setup to probe the magnetization states by measuring spin polarization, which can be feasible using current state-of-the-art STM techniques\cite{Mohr-noise-19,Mohr2021}. Surprisingly, the in-plane quantum transport for the B-G, which is non-spin polarized, can be spin-polarized by depositing FeTPP with a TSP of more than $10\%$ for typical coverages near the Fermi energy. This large electrically controlled TSP can be detected, for instance, in a spin valve setup \cite{Coronado20}. These results open an attractive route for the design of full-electrical writing and reading techniques in molecule/2D materials heterostructures.
	
	\noindent{\bf Acknowledgment}
	We acknowledge funding from the EU Horizon 2020 programme (Grant No. 766726) and the Danish National Research Foundation (Grant No. DNRF103). 
	\bibliographystyle{apsrev}
	\bibliography{References}
	
\end{document}